\documentclass[aps,prl,twocolumn,groupedaddress]{revtex4-1}
\usepackage{graphics}

\begin{document}
\newcommand{\tbti}{Tb$_2$Ti$_2$O$_7$}
\newcommand{\tb}{Tb$^{3+}$}
\newcommand{\mub}{$\mu_{\rm B}$}

\title{Tb$_2$Ti$_2$O$_7$, a two singlet spin-liquid pyrochlore}

\author{P.Bonville}
\email[]{pierre.bonville@cea.fr}
\affiliation{CEA, Centre de Saclay, DSM/IRAMIS/Service de Physique de l'Etat
Condens\'e, 91191 Gif-sur-Yvette, France}
\author{I.Mirebeau, A. Gukasov, S. Petit, J. Robert}
\affiliation{CEA, Centre de Saclay, DSM/IRAMIS/Laboratoire L\'eon Brillouin,
91191 Gif-sur-Yvette, France}

\date{\today}

\begin{abstract}
The pyrochlore compound \tbti\ is known to remain in a spin-liquid state down to the lowest attainable temperature (0.05\,K), whereas current theories predict it should order into an antiferromagnetic structure. A number of models have been developed in order to resolve this discrepancy, but they could not obtain a spin-liquid ground state. We present here an explanation for the lack of magnetic order in \tbti\ based on the presence of a Jahn-Teller-like distortion from the local trigonal symmetry and on the physics of the two-singlet system coupled by exchange.
\end{abstract}


\maketitle

Frustration in materials crystallizing in the pyrochlore lattice has been a subject
of interest since the pioneering work of J.Villain \cite{villain}, and an intense
burst of activity in the field has taken place in the last ten years (for a review,
see Ref.\cite{revging}). Among the various compounds, the titanates R$_2$Ti$_2$O$_7$
(R=rare earth) occupy a special position since many different ground states are
observed for the different R$^{3+}$ ions and an understanding of their basic properties
is now achieved \cite{revging}. The Tb titanate \tbti\ stands as an exception, since
its most salient property, i.e. the absence of magnetic ordering of the \tb\ moments
remains unexplained. Pioneering neutron and muon measurements \cite{gardner99} evidenced a cooperative paramagnetic state, where short range correlated Tb moments keep fluctuating down to at least 0.05\,K, failing to develop long range order (LRO) in spite of a high Curie-Weiss temperature ($\sim -$14~K) indicative of antiferromagnetic (AF) interactions. Since its discovery, \tbti\ has remained a theoretical puzzle as well as an experimental challenge. It is considered as the best example of a spin liquid (SL) \cite{balents10} in the pyrochlore lattice, in which quantum fluctuations are at play through a peculiar crystal electric field (CEF) scheme of the \tb\ ion. Indeed, it shows a low lying energy level (a doublet at $\Delta \simeq 18$\,K above the ground doublet) \cite{aleks,ging01,mirb07}, in contrast with classical Ising spin ices with Ho or Dy \cite{bram} where $\Delta \simeq 250-300$\,K. The paramagnetic spin correlations were understood taking into account these two low lying doublets \cite{kao}, and low energy spin fluctuations were interpreted by a dynamically induced frustration involving virtual crystal field excitations, leading to the concept of $"$quantum spin ice$"$ \cite{molavian07}. Indeed, local spin ice like configurations have been observed under applied field \cite{cao08,sazonov10}. In spite of all these attempts, the SL ground state of \tbti\ remains unexplained so far.

The \tb\ ion is a non-Kramers (or Van Vleck) ion, i.e. it possesses an even number (8)
of electrons in its $4f$ shell. Its ground spin-orbit state has a total momentum J=6 and a Land\'e factor $g_{\rm J}$=3/2.
A specific property of \tbti\ is that a strong $4f$ electron-strain interaction combined with its peculiar CEF level structure should lead to a {\it cooperative} Jahn-Teller (JT) transition at $T_{\rm JT} \simeq$0.1\,K \cite{mams}, which consists in a spontaneous symmetry lowering. Recently, high resolution x-ray scattering measurements down to 0.3\,K \cite{ruff07} have evidenced structural fluctuations below 15\,K, which are thought to be precursor to such a Jahn-Teller transition and which would occur between cubic and tetragonal symmetry.

We propose here that a cooperative tetragonal distortion develops in \tbti\ at low temperature, which breaks the local trigonal symmetry ($D_{3d}$) at the rare earth site. The degeneracy of the \tb\ CEF doublets is then lifted, and the low lying CEF states are two singlets separated by a small energy gap. For non-Kramers rare earths, the singlets are ``non-magnetic'', i.e. $\langle J_\alpha \rangle$=0 ($\alpha=x,y,z$), in the absence of a magnetic/exchange field. Exchange is nevertheless present at low temperature and tends to establish a magnetic ground state (and thus a magnetic LRO) by mixing with excited states. This competition between crystal field and exchange can result in a paramagnetic ground state if the exchange is not strong enough \cite{bleaney}. In other words, there is a critical value of interionic exchange below which the material does not show magnetic LRO down to T=0. Short range and dynamic spin correlations are present in this phase, which can be called a spin-liquid. We believe this mechanism is at play in \tbti\ and explains the absence of LRO. In this work, we apply this idea to explore in detail the consequences  of a distortion from trigonal symmetry on the ground state of \tbti.
This distortion probably occurs over a finite distance and during a finite lifetime at low temperature and preserves the overall cubic symmetry of the lattice.
For zero or small distortion, we show that AF exchange would lead to a magnetic LRO ground state, but that above a threshold distortion value, there appears a range of exchange constants where no LRO can settle in, i.e. the system remains in a spin liquid state down to $T$=0. We also show that the assumption of isotropic exchange is not sufficient, and that the introduction of anisotropic exchange, known to be present in the pyrochlore titanates \cite{cao09}, allows to quantitatively reproduce most properties of \tbti.

We start by computing the phase diagram as a function of exchange. We consider the total hamiltonian for a \tb\ ion: ${\cal H} = {\cal H}_{\rm CEF} + {\cal H}_{\rm ex} + {\cal H}_{\rm dip}$, containing the CEF interaction as in Ref.\cite{mirb07}, the short range exchange interaction with the 6 nearest neighbours and the infinite range dipolar interaction calculated using Ewald summation method \cite{ewald}. The ground state at a given temperature is obtained by a self-consistent mean field calculation involving the 4 sites on a tetrahedron. Then, only magnetic structures having {\bf k}=0 propagation vector can be reached. The exchange interaction is described in terms of a molecular field tensor $\tilde \lambda$, with ${\cal J} = (\frac{g_{\rm J}} {g_{\rm J}-1})^2\ \lambda$ \mub$^2$, where ${\cal J}$ is the usual exchange integral, and the molecular field acting on a given ion is: {\bf H}$_{\rm mol} = \tilde \lambda \sum_{k=1,6}${\bf m}$_{\rm k}$.
\begin{figure}
\resizebox{0.45\textwidth}{!}
{\includegraphics{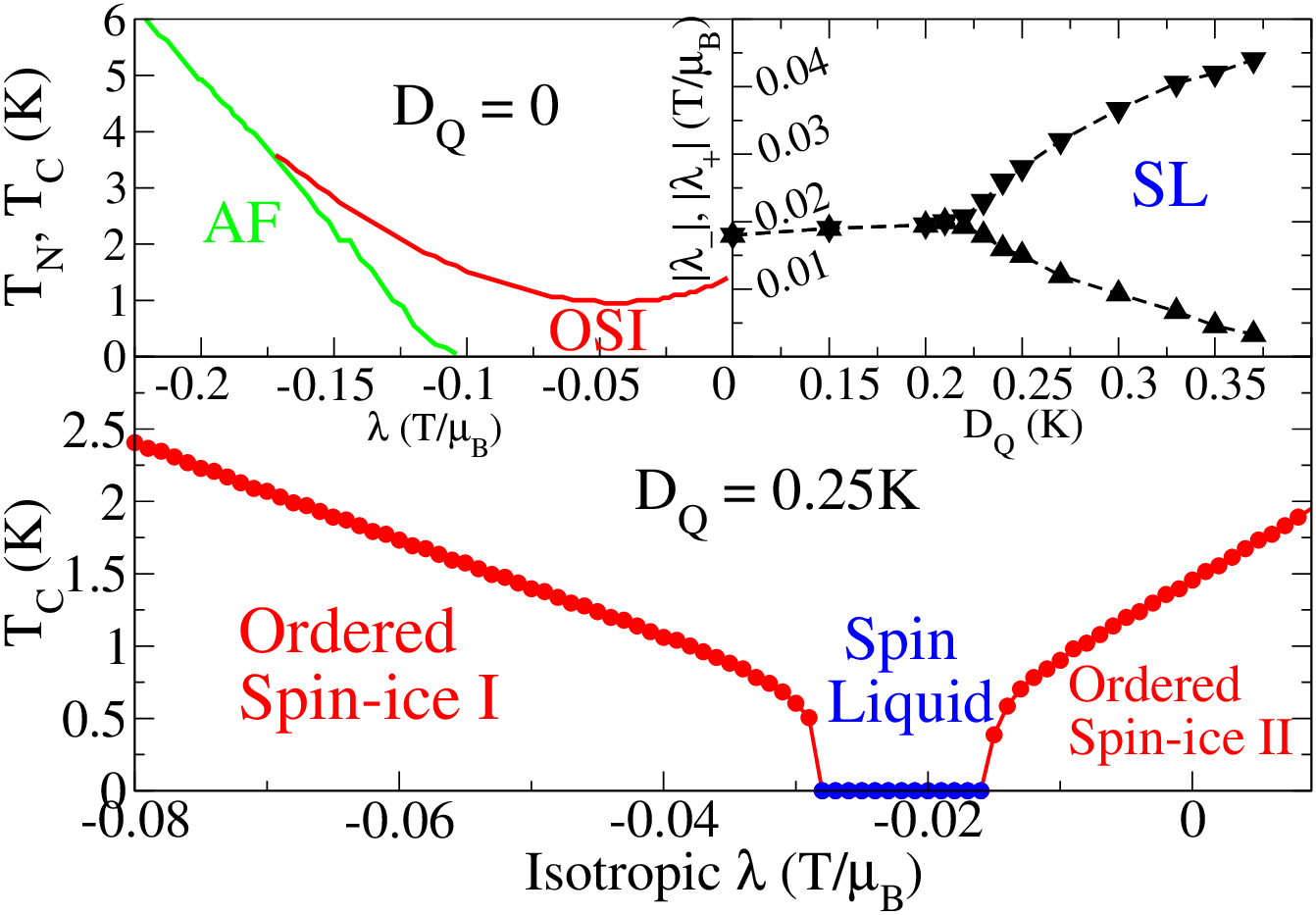}}
\caption{(T,$\lambda$) phase diagrams with the CEF level scheme of \tbti, with isotropic exchange and dipolar interaction. {\bf Upper left panel}: no distortion; {\bf lower panel}: with a tetragonal distortion $D_Q$=0.25\,K; {\bf upper right panel}: SL phase limits as a function of the distortion stength $D_Q$. The transition line to the AF phase is in green, to the OSI phase in red, and the SL phase is in blue.}
\label{diagp}
\end{figure}
We first set zero distortion (upper left panel in Fig.\ref{diagp}). Magnetic ordering is found to occur for all $\lambda$ values. We find there is a crossover value: $\lambda_{cr} = -0.10$\,T/\mub\ below which the ground state is an AF state (``all in/ all out''), and above which it is an ordered spin-ice phase (OSI, ``two in - two out'') akin to that observed in Tb$_2$Sn$_2$O$_7$ \cite{mirb05}. In the OSI phase, the Tb moments are slightly tilted with respect to their local [111] axis.
On decreasing $\vert \lambda \vert$ towards the ferromagnetic region, they tend towards [001], recovering results similar to those obtained in Ref.\cite{champ}. As in  the spin ice phase diagram of Ref.\cite{hertog}, we obtain the OSI phase for not too large {\it antiferromagnetic} exchange constants.

We next introduce a tetragonal distortion ${\cal H}_Q=D_Q J_Z^2$, where OZ is the cubic [001] axis. In the local frame with [111] as z-axis, the distortion term writes:
\begin{equation}
 {\cal H}_Q = \frac{D_Q}{3}\ [2J_x^2+J_z^2+\sqrt{2}(J_xJ_z+J_zJ_x)].
\label{dq}
\end{equation}
An interesting behaviour appears in the phase diagram as one switches on the distortion: there opens a gap in the low $\vert \lambda \vert$ region of the OSI transition line for $D_Q >$ 0.22\,K. The lower panel of Fig.\ref{diagp} shows this behaviour for $D_Q$=0.25\,K, which is the strength of the distortion in \tbti, as will be shown below. For $\lambda$ values inside the gap, the spontaneous moment at each \tb\ site is zero (down to 0.05\,K). This region corresponds to a phase with no long range order down to very low temperature and is analogous to that found in Ref.\cite{bleaney}. It should be the basic reason for the absence of LRO in \tbti. The two OSI phases obtained on either side of the gap are slightly different: phase OSI I is similar to the OSI phase obtained with zero distortion, whereas phase OSI II has a different symmetry. The gap width increases as $D_Q$ increases above 0.22\,K, as illustrated in the upper right panel of Fig.\ref{diagp}, where the lower and higher limits of the SL phase $\vert \lambda_- \vert$ and $\vert \lambda_+ \vert$ are plotted versus $D_Q$.

The distortion induces important changes in the eigen-functions of the ground states. In the basis $\vert m \rangle$=$\vert J=6; J_z=m \rangle$, the ground trigonal wave-functions are:
\begin{equation}
 \vert \psi_1 \rangle \simeq a \vert -4 \rangle+b \vert 5 \rangle\ {\rm and}\
 \vert \psi_2 \rangle \simeq -b \vert -5 \rangle+a \vert 4 \rangle\,
\label{fptrig}
\end{equation}
with $a \simeq$0.96 and $b \simeq$0.23. This non-Kramers doublet is such that the transverse matrix element $\langle \psi_2 \vert J_\alpha \vert \psi_1 \rangle$ is zero in zero magnetic field and remains close to zero as a moderate field is applied. Introducing the distortion (\ref{dq}) yields new eigenstates which are entangled singlets of symmetric/antisymmetric type:
\begin{equation}
\vert \psi_{s,a} \rangle = \frac{1}{\sqrt{2}} \ [ \vert \psi_1 \rangle \pm \vert \psi_2 \rangle] \simeq \frac{1}{\sqrt{2}}\ [\vert +4 \rangle \pm \vert -4 \rangle].
\label{asym}
\end{equation}
A particular property arises in \tbti\ from the structure of states (\ref{fptrig}): the tetragonal distortion lifts their degeneracy at first perturbation order, giving an energy separation of the two ground singlets $\delta = 12abD_Q \sqrt{11}$. For $D_Q$=0.25\,K, this yields $\delta$=2.2\,K=0.046\,THz. This property also holds in Tb$_2$Sn$_2$O$_7$, but not in Ho$_2$Ti$_2$O$_7$, for instance, where the splitting of the singlets for $D_Q$=0.25\,K would be a few mK. By contrast with the trigonal states (\ref{fptrig}), the transverse matrix element $\langle \psi_s \vert J_z \vert \psi_a \rangle$ is non-vanishing and large.

\begin{figure}
\resizebox{0.42\textwidth}{!}
{\includegraphics{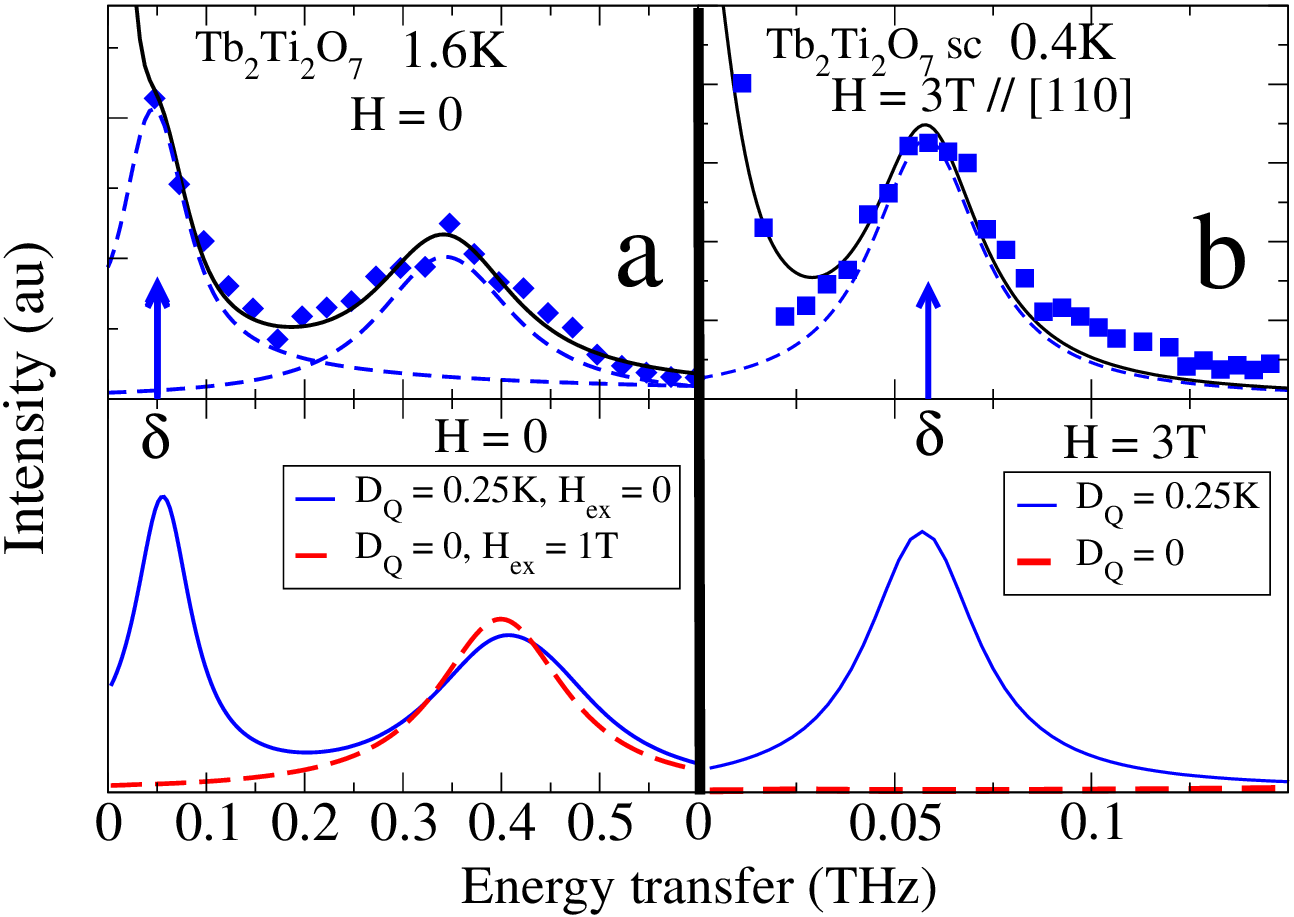}}
\caption{Neutron inelastic spectra in \tbti. The upper panels show the data, with dashed  Lorentzian-shaped lines marking the peak positions. The lower panels show the simulations with the full hamiltonian. {\bf a:} spectra in zero field at $T$=1.6\,K; data from Ref.\cite{mirb07} and simulations with a tetragonal distortion $D_Q$ = 0.25\,K (blue solid line) or with a Zeeman interaction with $H_{\rm ex}$=1\,T (red dashed line).  {\bf b:} in-field spectra with $H$=3\,T applied along [110], at $T$=0.4\,K; single crystal data from Ref.\cite{rule09} and simulations with the distortion $D_Q$=0.25\,K (blue solid line) or without distortion (red dashed line).}
\label{inel1k6}
\end{figure}

The low temperature inelastic neutron spectra, both in zero field \cite{rule06,mirb07} and in finite field \cite{rule06,rule09}, show a low energy inelastic line which gives a direct evidence for the presence of the distortion and allows its strength to be determined. At 1.6\,K in zero field, this line is not well resolved from the quasi-elastic line, but it is visible at a low energy $\delta \simeq 0.04$\,THz besides the CEF transition near 0.4\,THz (see Fig.\ref{inel1k6} {\bf a}). Its intensity decays more rapidly at large Q than the square of the form factor, demonstrating its magnetic origin.
At 0.4\,K in an applied field of 3\,T, the quasi-elastic line is much narrower and the inelastic line, which has slightly shifted to an energy $\delta \simeq$0.055\,THz, is very well resolved (see Fig.\ref{inel1k6} {\bf b}). This line arises from a degeneracy lifting of the ground doublet, and the simulations in the lower panel of Fig.\ref{inel1k6} demonstrate that the tetragonal distortion alone can generate this degeneracy lifting and account for a sizeable intensity transition (blue solid line). Both the hypothetical presence of a static exchange field of 1\,T for the zero field spectrum, or the presence of the applied field alone for the 3\,T spectrum result in a quasi-extinct line (red dashed lines in Fig.\ref{inel1k6}). These extinction rules originate in the above considerations about the wave-functions, the intensity $I_{\rm ij}$ of a transition between states $\vert i \rangle$ and $\vert j \rangle$ being proportional to $\sum_{\alpha=x,y,z} \vert \langle i \vert J_\alpha \vert j \rangle \vert^2$. From the position of this line, the strength of the tetragonal distortion in \tbti\ is estimated to be $D_Q$=0.25\,K.

The low energy part of the CEF spectrum in the presence of the distortion consists in 4 non-degenerate states, i.e. 2 sets of closely spaced singlets with a splitting of about 2\,K, separated by 18\,K. Therefore, specific heat measurements carried out from a temperature lower than the splitting of the ground singlets should derive an entropy release $\Delta S$ close to $R \ln 4$ as soon as the 4 states are populated, instead of $R \ln 2$ for 2 degenerate doublets. Specific heat data in \tbti\ \cite{hama,chapuis} have indeed obtained $\Delta S \simeq R \ln 4$ between 0.15 and 20\,K, confirming the level structure proposed by our model.

One must now examine whether the other available experimental data in \tbti\ are compatible with $\lambda$ values within the SL gap associated with $D_Q$=0.25\,K, i.e. $-0.028 \le \lambda \le -0.015$\,T/\mub. From the high temperature susceptibility data, an estimation was given: $\lambda \simeq -0.05$\,T/\mub\ \cite{mirb07}, not far from, but outside the SL range. However, other data (the thermal variation of the local susceptibility \cite{cao09} and the field evolution of the induced magnetic structure \cite{sazonov10}) point to the presence of {\it anisotropic exchange}.
\begin{figure}
\resizebox{0.44\textwidth}{!}
{\includegraphics{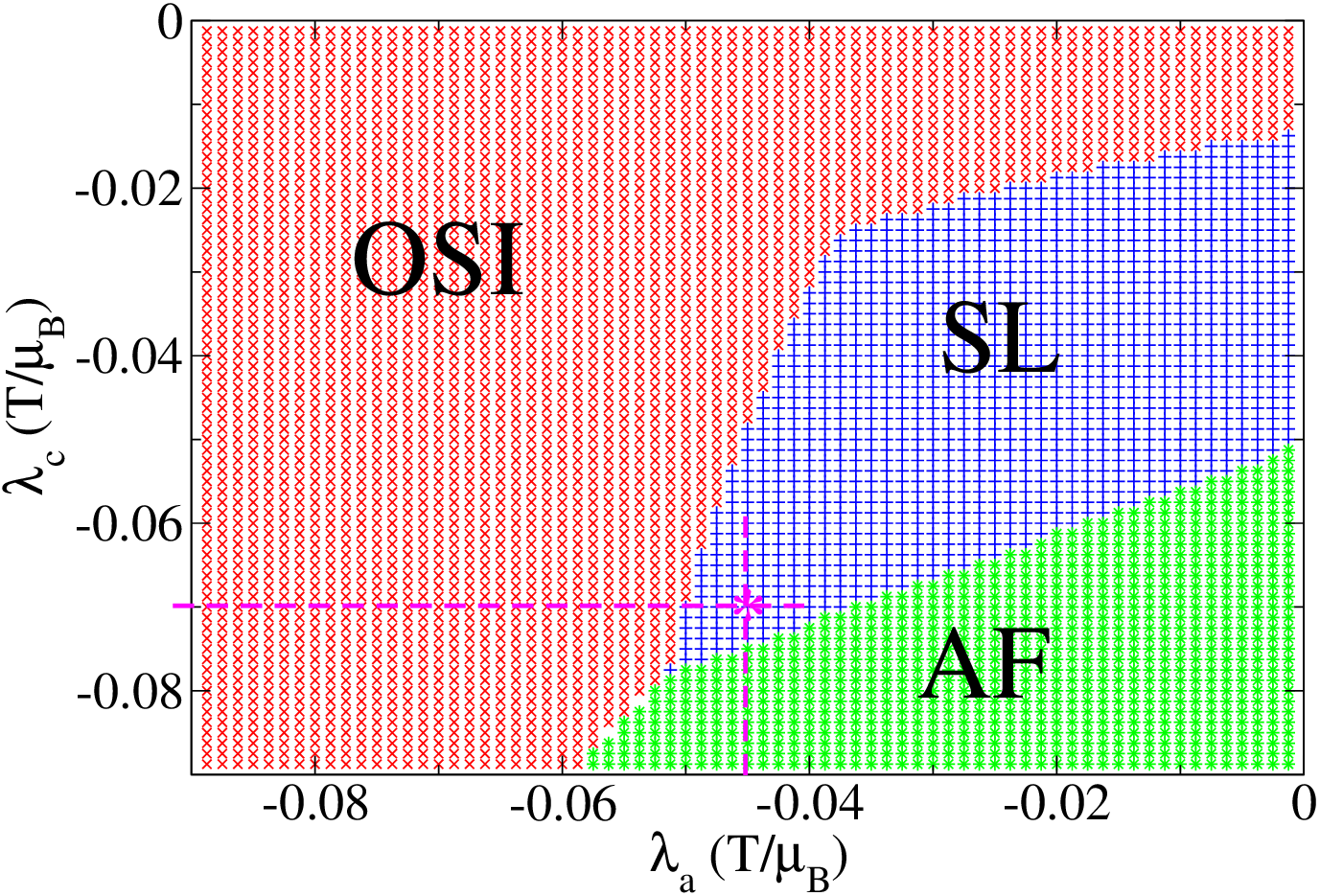}}
\caption{($\lambda_c,\lambda_a$) phase diagram with the CEF level scheme of \tbti\ at $T$=0 in the presence of anisotropic exchange, dipolar interaction and a tetragonal distortion $D_Q$=0.25\,K, with the fixed value $\lambda_b=-0.13$\,T/\mub. The red region has the OSI phase as ground state, the blue region the SL phase and the green region the AF phase. The point $\tilde \lambda_0$ =\{$-0.045,-0.13,-0.07$\}\,T/\mub, marked with a pink star, could correspond to \tbti\ (see text).}
\label{diagt0}
\end{figure}
\begin{figure}
\resizebox{0.42\textwidth}{!}
{\includegraphics{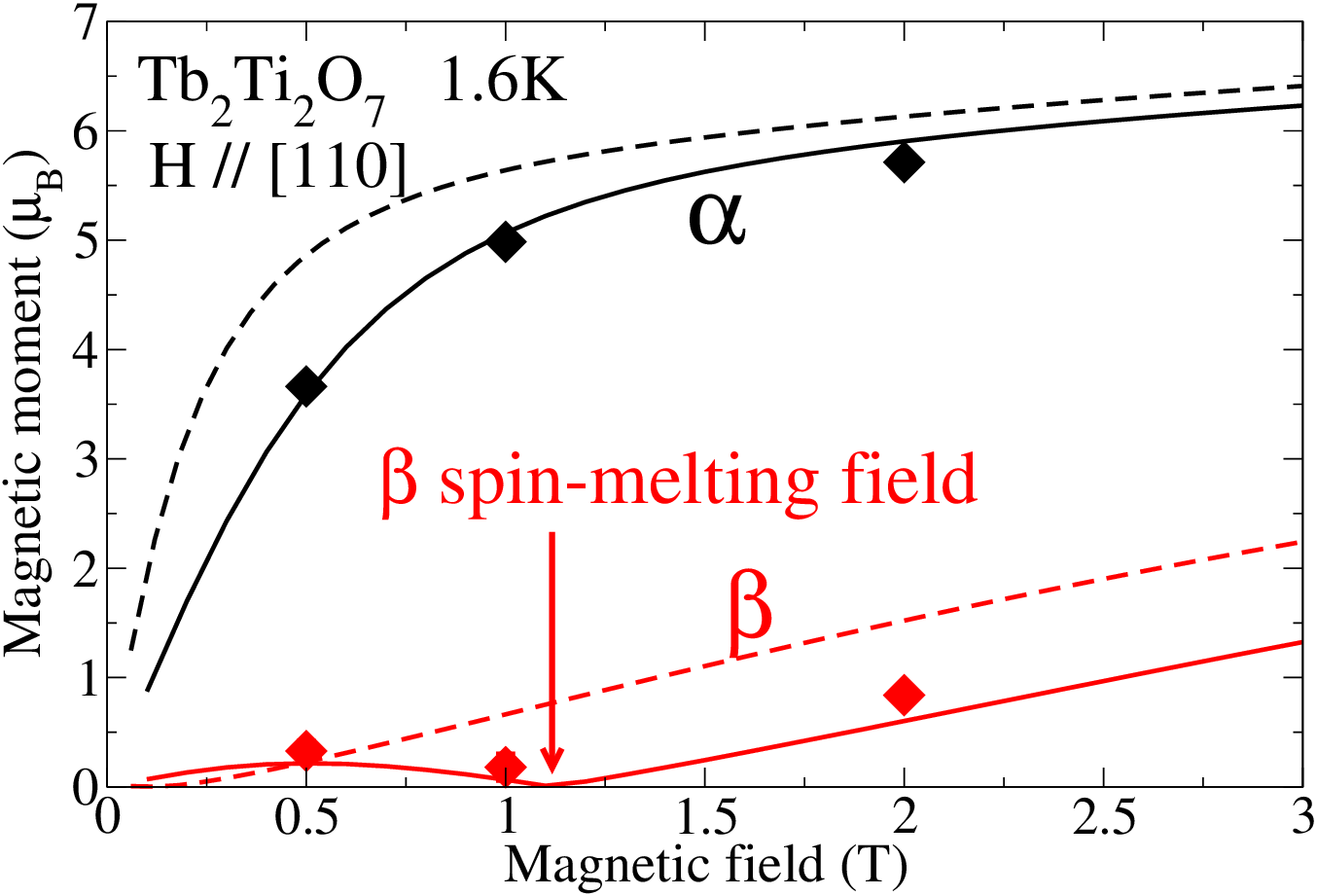}}
\caption{Calculations of the field variations, for {\bf H} // [110], of the $\alpha$ and $\beta$ \tb\ moments at 1.6\,K, showing the spin-melting field for $\beta$ moments, obtained with the molecular field tensor $\tilde \lambda_0$ (see Fig.\ref{diagt0}) (solid lines). The dashed lines represent the calculation with the isotropic molecular field constant $\lambda=-$0.025\,T/\mub. The data are from Ref.\cite{sazonov10}.}
\label{spmelt}
\end{figure}
So we have determined the phase diagram at $T$=0 assuming a 3-component molecular field tensor \{$\lambda_a$,$\lambda_b$,$\lambda_c$\}, where the {\bf c} axis is chosen along the bond linking two Tb ions. A representative cut of this phase diagram, for $\lambda_b=-$0.13\,T/\mub, is shown in Fig.\ref{diagt0}. For large $\vert \lambda \vert$ values, the OSI phase is the ground state (in red), but a sizeable SL region (in blue) together with an AF region (in green) are found for lower $\vert \lambda \vert$ values.

A quantity which is rather sensitive to the $\lambda$ values is the in-field behaviour of the magnetic structure \cite{sazonov10}. In particular, the $\beta$-moments (those having their trigonal axis perpendicular to the field {\bf H} applied along [110]) undergo a ``spin-melting'' at $H_{\rm m}$ = 1.5(5)\,T, where their magnitude vanishes, as shown in Fig.\ref{spmelt}. The curves computed using the anisotropic $\tilde \lambda_0$ tensor in the SL region of Fig.\ref{diagt0} (solid lines) reproduce the data better than those obtained using the isotropic value $\lambda=-0.025$\,T/\mub\ (dashed lines), corresponding to a point in the SL range of Fig.\ref{diagp}.
\begin{figure}
\resizebox{0.42\textwidth}{!}
{\includegraphics{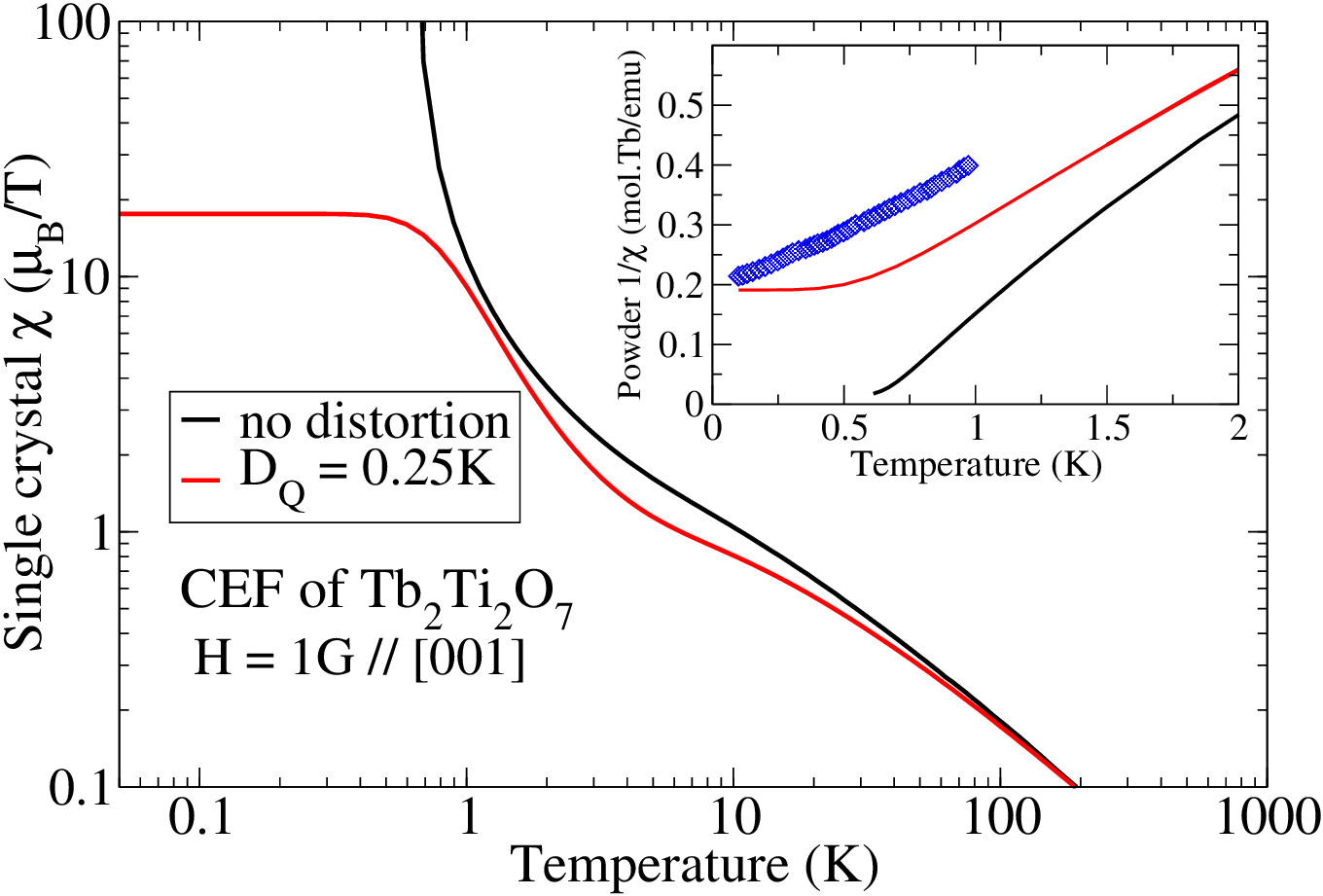}}
\caption{Thermal variation of the single crystal magnetic susceptibility calculated with the CEF parameters for \tbti, with the anisotropic $\tilde \lambda_0$ tensor derived above, in the absence and in the presence of the tetragonal distortion. Insert: inverse powder susceptibilities with the same assumptions and legend, and experimental data from Ref.\cite{luo} (blue squares).}
\label{ximo}
\end{figure}

The magnetic susceptibility $\chi(T)$ at very low temperature should reflect the presence of the distortion, as show the calculated curves in Fig.\ref{ximo}: for $D_Q$=0 (black curves), $\chi(T)$ shows a Curie-Weiss behaviour, whilst for $D_Q$=0.25\,K (red curves) it shows a Van Vleck-like behaviour with a saturation below 1\,K, as expected for a singlet ground state separated from the excited state by about 2\,K. The powder susceptibility measured by magnetometry \cite{luo} does not saturate (see insert of Fig.\ref{ximo}, blue squares), but $1/\chi$ computed without distortion is farther from experimental data than that computed with distortion. The powder susceptibility can also be measured through the $\mu$SR Knight shift in a transverse field. The Knight-shift data of Ref.\cite{dunsi} do not evidence any net saturation, but those in Ref.\cite{ofer} do show a clear saturation below 1-2\,K, in good agreement with our model.

The numerous neutron scattering studies performed in \tbti\ \cite{gardner99,gardner01,yasui} have shown that the \tb\ ions behave like short range correlated moments down to low temperature. Our picture is fully compatible with fluctuating Tb moment components with sizeable values. The fast spin fluctuation rate down to the 0.05\,K range derived by $\mu$SR \cite{gardner99} and neutron spin echo \cite{gard03} can be explained by exchange/dipole driven relaxation between \tb\ singlets of type (\ref{asym}), which are linked by large matrix elements of $J_z$. Indeed, according to the Fermi rule, the relaxation rate is proportional to $\vert \langle \psi_a \vert J_z \vert \psi_s \rangle \vert^2$. Below about 0.5\,K, the spin fluctuations freeze at the 10$^{-11}$\,s time scale \cite{yasui}, and coexist with spin-glass irreversibilities \cite{luo} and with magnetic correlations extending over a few cubic cells, with the same propagation vector {\bf k}=(001) as the LRO induced under stress \cite{mirb04}.

As a conclusion, we think the absence of magnetic LRO in \tbti\ is caused by the development at low temperature of a precursor Jahn-Teller distortion which makes this material a two-singlet system, where AF exchange is not strong enough to allow for magnetic ordering. Taking also into account exchange anisotropy, most of the properties in \tbti\ can be explained. The in-field spin-wave spectrum in \tbti\ \cite{rule06} will be accounted for in a future work. Our model can also be applied successfully to explain the characteristics of the OSI phase in Tb$_2$Sn$_2$O$_7$ \cite{mirb05}; this will be the subject of a future publication.


\begin{thebibliography}{99}
\bibitem{mirb07}
I. Mirebeau, P. Bonville, M. Hennion, Phys.Rev. B {\bf 76}, 184436 (2007) (the $\lambda$ values in this work must be divided by 6 for comparison with the present work.)
\end{thebibliography}
\end{document}